\documentclass[12pt,a4paper]{article}
\usepackage{amsmath,amssymb,bm,ascmac,bbm,mathtools}
\usepackage[dvipdfmx, usenames]{color}
\usepackage[dvipdfmx]{graphicx}
\usepackage{xcolor}
\usepackage{here}
\usepackage{authblk}

\newcommand{\lessim}{\mathop{<}\limits_{\displaystyle{\sim}}}

\newcommand{\tr}{\text{tr}}

\newcommand{\mcal}{\mathcal}
\newcommand{\mbb}{\mathbb}

\newcommand{\ra}{\rangle}

\begin{document}
\title{Emergent symmetry and free energy}
\author{Ken KIKUCHI}
\affil{Yau Mathematical Sciences Center,
Tsinghua University}
\date{}
\maketitle

\begin{abstract}
Just as spontaneous symmetry breaking can be understood in terms of energy, emergent symmetry (more precisely, its `size' and structure) can also be explained by free energy. In particular, in renormalization group flow to rational conformal field theory, we find infrared symmetry category is realized by consistent modular tensor category with minimal free energy. For demonstration, we study non-unitary theories.
\end{abstract}


\makeatletter
\renewcommand{\theequation}
{\arabic{section}.\arabic{equation}}
\@addtoreset{equation}{section}
\makeatother

\section{Introduction and summary}
Spontaneous symmetry breaking (SSB) can be understood from the viewpoint of energy. The free energy $F$ is given by energy $E$, temperature $T$, and entropy $S$ as
\[ F=E-TS. \]
At high temperature, the second term is dominant. Thus, states with large entropy are realized to minimize the free energy. This typically breaks order. On the other hand, at low temperature, the second term is unimportant, and small $E$ states are realized at the cost of entropy. This typically realizes ordered states. For instance, let us recall ferromagnet. At high temperature, the entropy term is dominant. Hence, ferromagnets try to minimize the free energy by orienting spins randomly and making the entropy large. In this state, the rotation symmetry is preserved. On the other hand, at low temperature, the entropy term is negligible, and ferromagnets try to minimize the free energy by aligning spins. This gives an order, and breaks the rotation symmetry spontaneously. The main goal of this paper is to explain (some aspects of) emergent symmetry from this perspective.

Emergent symmetry is a symmetry which is absent in ultraviolet (UV) and appear in infrared (IR). This phenomenon is called symmetry enhancement. In particular, emergent symmetries are common in two dimensional massless renormalization group (RG) flows. Recently, the underlying mechanism behind the phenomenon has been understood focusing on RG flows to rational conformal field theories (RCFTs) \cite{K21,K22}. The reason why emergent symmetries should appear is the following. RCFTs are mathematically described by modular tensor categories (MTCs) \cite{MS1,MS2}. This is a braided fusion category (BFC)\footnote{In modern language, symmetries are generated by topological operators supported on defects with codimension $(q+1)$ \cite{GKSW,BT17}. Symmetries generated by the topological operators are called $q$-form symmetries. They in general do not have inverse elements. Such symmetries are called non-invertible. The symmetries, which fail to form a group, are in general described by certain categories. The category is called symmetry category. Just like group multiplication, objects of symmetry category can be fused to form other objects. Thus, they are in general given by fusion categories equipped with fusion. If a fusion category $\mcal C$ admits a braiding $c$, a pair $(\mcal C,c)$ is called braided fusion category (BFC).} with an invertible (topological) $S$-matrix. (Such BFCs are called modular.) When one performs relevant deformation to a theory with BFC, some objects are broken, while the others survive all along the RG flow. The surviving topological defect operators again form braided fusion subcategory. In general, the subcategory does not admit invertible (topological) $S$-matrix. However, if the IR theory is an RCFT, the symmetry category should be modular, and it is \textit{forced} to have invertible (topological) $S$-matrix. Therefore, emergent topological defect operators appear at conformal fixed points, and make BFCs modular. This is one reason why symmetries emerge. Even when a surviving BFC is modular, it may not admit central charge smaller than that in UV. If that is the case, in unitary theories, the surviving MTC should be enlarged to another MTC with smaller central charge to make it compatible with the $c$-theorem \cite{cthm}. This is another reason why symmetries emerge. In short, in rational massless RG flows, symmetry categories in IR should be both modular and consistent with the $c$-theorem.

How the emergent symmetries can be understood in terms of free energy? Topological entanglement entropy gives an answer. Kitaev and Preskill \cite{KP05} found that an MTC gives universal contribution to entropy (see also \cite{LW05}). More concretely, the contribution is given as follows. Each object $j$ of an MTC $\mcal C$ has specific number $d_j$ called quantum dimension. (In unitary theories, we have $d_j\ge1$.) Using the quantum dimensions, we can define global (quantum) dimension $D$ of the category $\mcal C$ via
\begin{equation}
    D^2:=\sum_{j\in\mcal C}d_j^2.\label{globaldim}
\end{equation}
(In unitary theories, $D$ is positive.) Then, the topological entanglement entropy is given by
\begin{equation}
    S\ni-\gamma=-\ln D.\label{TEE}
\end{equation}
The contribution enters free energy as
\begin{equation}
    F\ni T\ln D.\label{FD}
\end{equation}
(In their argument, the temperature $T$ is introduced as length $1/T$ of the Euclidean time compactified to a circle.) The formula makes clear that MTCs with smaller global dimensions are energetically favored. In other words, the larger global dimensions are, the higher free energy becomes. This observation explains and sharpens ``naturalness'' suggested in \cite{K21}; a consistent symmetry category with the smallest global dimension should be realized. If symmetry categories are enlarged at the cost of free energy, there should be reasons. In examples studied in \cite{K21,K22}, namely RG flows to RCFTs, the reasons are modularity or consistency with the $c$-theorem.

Armed with this new insight provided by free energy, we can answer various questions on emergent symmetries: when they appear, how large they are, and what are their structures. We will demonstrate this in examples below. In particular, given classifications of MTCs, we will see that we can often fix which MTC is realized in IR.

For demonstration, we choose non-unitary RCFTs. There are two reasons for our choice. Firstly, our understanding of symmetry enhancement so far heavily relies on unitarity (through the requirement of $c$-theorem). Thus, one may think our explanation would not work if we relax the assumption of unitarity. Addressing this concern is our first reason why we study non-unitary theories. The second reason is the rank\footnote{The `size' of category is called rank. More precisely, it counts the number of isomorphism classes of simple objects.} of MTCs. Typically, non-unitary RCFTs have smaller ranks. (For example, the three lowest unitary CFTs are described by MTCs with ranks 3, 6, 10, while those of non-unitary CFTs have ranks 2, 3, 4.) Therefore, we do not need classifications of higher rank MTCs. In fact, we will see that IR MTCs can be completely fixed up to our third examples.

How can we explain symmetry enhancement in non-unitary RCFTs? We simply replace the $c$-theorem with the $c^\text{eff}$-theorem \cite{CDR17}. The theorem claims
\begin{equation}
    0\le c^\text{eff}_\text{IR}\le c^\text{eff}_\text{UV}.\label{ceffthm}
\end{equation}
The effective central charge is defined by
\begin{equation}
    c^\text{eff}:=c-24\Delta_\text{smallest},\label{ceff}
\end{equation}
where $\Delta_\text{smallest}$ is the smallest conformal dimension in the theory. In unitary theories, we have $h\ge0$. Thus the smallest conformal dimension is always provided by the identity operator, $\Delta_\text{smallest}=0$, and the original $c$-theorem is recovered from the $c^\text{eff}$-theorem. What is nice about the theorem for our purposes is that the $c^\text{eff}$-theorem generalizes the $c$-theorem, and works even in non-unitary theories. For the theorem to work, however, we have to assume unbroken $PT$ symmetry.\footnote{A $CPT$-symmetric theory with trivial charge conjugation is automatically $PT$-symmetric.} More explicitly, since $PT$ symmetry in IR is nontrivial, we \textit{impose} the $c^\text{eff}$-theorem. (Namely, we require modularity and consistency with the $c^\text{eff}$-theorem.)

In order to answer questions on emergent symmetry listed above (and check our answers), we study known RG flows to non-unitary (bosonic) RCFTs. We try to find consistent MTC with the smallest global dimension making the most of constraints on RG flows. The constraints we employ are summarized as follows:
\begin{itemize}
    \item simplicity of surviving topological defects,
    \item fusion ring,
    \item $F$-symbols,
    \item spin constraint,
    \item double braiding relation,
    \item and ``monotonicity'' of scaling dimensions.
\end{itemize}
Let us briefly explain each item. (For review/explanation of these constraints, see \cite{K22}.) We will focus on zero-form symmetries in two dimensions. Thus, they are generated by topological defect lines (TDLs). If a TDL commutes with relevant operators, it survives all along the RG flow triggered by the operators \cite{G12,CLSWY}. The simplicity of surviving TDLs in particular means the number cannot decrease. Thus, if the surviving TDLs do not form a consistent symmetry category, the only way out is to \textit{increase} the number with emergent TDLs. If several TDLs are preserved, we can fuse them. Since all TDLs (including resulting TDLs) commute with relevant operators, the fusion ring is invariant under RG flows. The $F$-symbols associated to them are also invariant under the flows. (This also means anomaly matching.) The $F$-symbols give another constraint \cite{KCXC}; for a surviving line $j$, there is an associated defect Hilbert space $\mcal H_j$. The space has operators with specific spin contents $S_j$. If relevant operators are (spacetime) scalars, the deformation preserves rotation symmetry. Hence spin contents are conserved. More precisely, some operators may be lifted along the flow, so spin contents in IR should be a subset of that in UV:
\begin{equation}
    S_j^\text{IR}\subset S_j^\text{UV}.\label{spinconst}
\end{equation}
Relevant spin contents for our study are listed in Appendix \ref{spin}. The double braiding relation means the following. Let us pick two surviving TDLs $i,j$ ($i,j$ can be the same). Then their double braidings are the opposite in UV and IR:
\begin{equation}
    c_{j,i}^\text{IR}c_{i,j}^\text{IR}=\left(c_{j,i}^\text{UV}c_{i,j}^\text{UV}\right)^*.\label{dbrelation}
\end{equation}
As a corollary of the double braiding relation, we obtain relations between the (topological) $S$-matrices in UV and IR
\[ \left(\widetilde S_\text{top}^\text{IR}\right)_{ij}=\left(\widetilde S_\text{top}^\text{UV}\right)_{ij}^* \]
by taking the (quantum) trace
\begin{equation}
    \left(\widetilde S_\text{top}\right)_{ij}:=\tr(c_{j,i}c_{i,j}).\label{topS}
\end{equation}
This in particular implies matching of quantum dimensions
\begin{equation}
    d_j:=\left(\widetilde S_\text{top}\right)_{1j}\label{qdim}
\end{equation}
because they are real.\footnote{A proof is as follows. One definition of quantum dimensions is
\[ d_j:=\frac{(S_\text{top})_{1j}}{(S_\text{top})_{11}}, \]
where $S_\text{top}:=\widetilde S_\text{top}/D$ is the normalized topological $S$-matrix. Here, topological $S$-matrix obeys
\[ (S_\text{top})_{ij}=(S_\text{top})_{i^*j}^*. \]
Recalling the reality of the identity $1^*=1$, we get
\[ d_j^*=\frac{(S_\text{top})_{1j}^*}{(S_\text{top})_{11}^*}=\frac{(S_\text{top})_{1j}}{(S_\text{top})_{11}}=d_j. \]} Finally, the ``monotonicity.'' It was found that conformal dimensions decrease ``monotonically.'' More precise statement is the following. In diagonal RCFTs, there is a one-to-one correspondence between primaries and TDLs (especially called Verlinde lines). Thus, for a surviving Verlinde line $j$, one primary each correspond to it in UV and IR. Let us denote their conformal dimensions $h_j^\text{UV}$ and $h_j^\text{IR}$, respectively. We found they obey
\begin{equation}
    h_j^\text{IR}\le h_j^\text{UV},\label{monotonic}
\end{equation}
and proved this in case of bosonic unitary discrete series of minimal models \cite{K22}. We can also prove the ``monotonicity'' for RG flows among non-unitary RCFTs we study (see Appendix \ref{statements}). Therefore, the ``monotonicity'' is clearly not a consequence of unitarity, but seems a feature of RG flow itself.

We employ these constraints at our disposal. We further use one empirical fact:
\begin{equation}
    \frac25\le c^\text{eff}.\label{ceffbound}
\end{equation}
The equality is saturated by the Lee-Yang model $c^\text{eff}_\text{LY}=\frac25$, and all the other nontrivial (R)CFTs we know have larger effective central charges. We do not have a proof of this fact, nor do not know whether this is true. However, by \textit{assuming} this, we manage to fix IR symmetry categories when classifications of MTCs are available. Our success indicates that emergent symmetries appear to realize consistent symmetry category with the smallest free energy.

\section{Examples}
Examples we study are RG flows among non-unitary minimal models. The models are labeled with two coprime natural numbers $p,q$, and denoted $M(p,q)$. We always take $q>p+1$, and consider $p\ge3$. The model has central charge
\begin{equation}
    c=1-\frac{6(p-q)^2}{pq}.\label{cpq}
\end{equation}
The theory has $\frac{(p-1)(q-1)}2$ primary operators (and hence the same number of Verlinde lines) labeled with Kac indices
\begin{equation}
    E_{p,q}:=\left\{(r,s)|1\le r\le q-1\&1\le s\le p-1\right\}/\sim,\label{Kac}
\end{equation}
where the relation is given by
\begin{equation}
    (r,s)\sim(q-r,p-s).\label{Kacequiv}
\end{equation}
The conformal dimensions of the primaries are given by
\begin{equation}
    h_{r,s}=\frac{(pr-qs)^2-(p-q)^2}{4pq}.\label{confdim}
\end{equation}
The model has $S$-matrix (modular and topological $S$-matrices coincide)
\begin{equation}
    S_{(r,s),(r',s')}=(-1)^{1+rs'+sr'}\sqrt{\frac8{pq}}\sin\left(\pi\frac pqrr'\right)\sin\left(\pi\frac qpss'\right).\label{Smatrix}
\end{equation}
A Verlinde line $\mcal L_{r,s}$ acts on a primary $\phi_{t,u}$ as
\begin{equation}
    \hat{\mcal L}_{r,s}|\phi_{t,u}\ra=\frac{S_{(r,s),(t,u)}}{S_{(1,1),(t,u)}}|\phi_{t,u}\ra.\label{action}
\end{equation}

Given these data, we can explicitly work out constraints on RG flows. The massless flows we consider are
\begin{equation}
\begin{split}
    M(p,2p+1)+\phi_{5,1}&\to M(p,2p-1),\\
    M(p,2p-1)+\phi_{1,2}&\to M(p-1,2p-1).
\end{split}\label{flows}
\end{equation}
These flows were found and studied in \cite{Z90,Z91,M91,RST94,DDT00} using the thermodynamic Bethe ansatz approach. We can prove the surviving BFCs are always modular in these flows (see Appendix \ref{statements}).\footnote{A useful tool to judge modularity of BFCs is the monodromy charge matrix
\begin{equation}
    M_{ij}:=\frac{\left(S_\text{top}\right)_{ij}\left(S_\text{top}\right)_{11}}{\left(S_\text{top}\right)_{1i}\left(S_\text{top}\right)_{1j}}.\label{M}
\end{equation}
An object $i\in\mcal C$ is transparent iff $\forall j\in\mcal C,M_{ij}=1$. Equivalently, if such a nontrivial object exists, the BFC is not modular.} Thus our main task is to compute effective central charges. We start from UV theories with smaller ranks.

\begin{itemize}
    \item $M(3,5)+\phi_{1,2}$\\
    The UV theory has $c_\text{UV}=-\frac35,c_\text{UV}^\text{eff}=\frac35$. The $\phi_{1,2}$-deformation preserves two TDLs $\{1,\mcal L_{3,1}\}$. The non-invertible Fibonacci line $\mcal L_{3,1}$ is associated to a primary with conformal dimension $\frac15$. Their double braidings are thus given by
    \[ \begin{pmatrix}id_1&id_{3,1}\\id_{3,1}&e^{-\frac{4\pi i}5}id_1\oplus e^{-\frac{2\pi i}5}id_{3,1}\end{pmatrix}. \]
    Taking the quantum trace, we obtain the unnormalized topological $S$-matrix
    \[ \widetilde S_\text{top}=\begin{pmatrix}1&-\zeta^{-1}\\-\zeta^{-1}&-1\end{pmatrix}. \]
    Dividing with quantum dimensions, we get the monodromy charge matrix
    \[ M=\begin{pmatrix}1&1\\1&-\zeta^2\end{pmatrix}. \]
    One sees the symmetric centralizer is trivial. This means the rank two surviving BFC is actually an MTC. According to \cite{GK94} (see also \cite{rank4,rank5}), there is only one rank two MTC with the same fusion ring. It has $SU(2)_3/\mbb Z_2$ realization, and has central charge $c=\frac{4n+2}5$ with $n<5,n\neq2$. Can the rank two MTC describe IR CFT? To answer this question, let us predict IR conformal dimension associated to the surviving line $\mcal L_{3,1}\to j$.
    
    The double braiding relation requires $j$ to obey
    \[ c_{j,j}^\text{IR}c_{j,j}^\text{IR}=e^{\frac{4\pi i}5}id_1\oplus e^{\frac{2\pi i}5}id_j. \]
    The $j$-channel predicts
    \[ \theta_j^{-1}=e^{-2\pi ih_j^\text{IR}}=e^{\frac{2\pi i}5}, \]
    or
    \[ h_j^\text{IR}=\frac45\quad(\text{mod }1). \]
    The ``monotonicity'' gives an upper bound on $h^\text{IR}_j$; $h^\text{IR}_j\le\frac15$. The inequality gives $h^\text{IR}_j=-\frac15,-\frac65,-\frac{11}5,\dots$ . The new inequality $h^\text{IR}_j\le-\frac15$ implies $-24h^\text{IR}_j\ge\frac{24}5$. For $c^\text{eff}$ to be no larger than $c^\text{eff}_\text{UV}=\frac35$, we need
    \[ c_\text{IR}\le-\frac{21}5. \]
    We know the surviving rank two MTC can have such central charge (e.g. $n=-6$, or $c=-\frac{22}5$). Therefore, our proposal says symmetry enhancement is unnecessary, and indeed the known RG flow $M(3,5)\to M(2,5)$ is realized by the minimal choices $h^\text{IR}_j=-\frac15,c=-\frac{22}5$, or $c^\text{eff}_\text{IR}=\frac25$. The spin constraints are satisfied by the matching
    \begin{equation} 
    \begin{array}{ccc}
    \text{UV}:&1&\mcal L_{3,1}\\
    &\downarrow&\downarrow\\
    \text{IR}:&1&\mcal L_{3,1}
    \end{array}.\label{M35toM25}
    \end{equation}
    \item $M(3,7)+\phi_{5,1}$\\
    The UV theory has $c_\text{UV}=-\frac{25}7,c^\text{eff}_\text{UV}=\frac57$. The $\phi_{5,1}$-deformation preserves two TDLs $\{1,\mcal L_{1,2}=\eta\}$. The $\mbb Z_2$ line $\eta$ is associated to a primary with conformal dimension $\frac54$. The double braidings are given by
    \[ \begin{pmatrix}id_1&id_\eta\\id_\eta&-id_1\end{pmatrix}. \]
    Taking the quantum trace, we obtain the unnormalized topological $S$-matrix
    \[ \widetilde S_\text{top}=\begin{pmatrix}1&-1\\-1&-1\end{pmatrix}. \]
    Dividing with quantum dimensions, we get the monodromy charge matrix
    \[ M=\begin{pmatrix}1&1\\1&-1\end{pmatrix}. \]
    We find there is no nontrivial transparent line, and the surviving rank two BFC is actually an MTC. According to \cite{GK94}, there is only one rank two MTC with $\mbb Z_2$ fusion ring. It has $SU(2)_1$ realization, and central charge $c=1,3$ mod $4$. Thus our proposal says the rank two MTC does not have to enhance unless it is inconsistent with the $c^\text{eff}$-theorem.
    
    In order to answer whether the rank two MTC should be enhanced or not, let us study the IR conformal dimension associated to the surviving $\eta$ line. The double braiding relation requires $\eta\to j$ to obey
    \[ c_{j,j}^\text{IR}c_{j,j}^\text{IR}=-id_1. \]
    The identity-channel predicts
    \[ \theta_j^{-2}=e^{-4\pi ih_j^\text{IR}}=-1, \]
    or
    \[ h_j^\text{IR}=\frac14\quad(\text{mod }\frac12). \]
    The ``monotonicity'' $h^\text{IR}_j\le h^\text{UV}_\eta=\frac54$ gives $h^\text{IR}_j=\frac54,\frac34,\frac14,\dots$ . Note that the $\mbb Z_2$ anomaly is matched by all candidates because $[5]=[3]=[1]=\dots$ mod $2$. One notices that, for these IR conformal dimensions, the effective central charge $c^\text{eff}$ cannot be a non-negative number smaller than $c^\text{eff}_\text{UV}=\frac57$.\footnote{The easiest way to see this fact is as follows. Since both $c$ and $-24\Delta_\text{smallest}$ are integers, the smallest non-negative effective central charge one can get is $1$. This is not smaller than $c^\text{eff}_\text{UV}$. Here, one may ask why we ruled out $c^\text{eff}_\text{IR}=0$, but one can show this cannot be realized by the surviving rank two MTC. To show this, let us set $c=1+2n$ with $n\in\mbb Z$ and $h^\text{IR}_j=\frac{5-2m}4$ with $m\in\mbb N$. Since the central charge itself is odd, to realize $c^\text{eff}_\text{IR}=0$, we need $h^\text{IR}_j<0$. Then the effective central charge is given by
    \[ c^\text{eff}_\text{IR}=1+2n-24\frac{5-2m}4=2n+12m-29. \]
    This cannot be zero.\label{rulingoutr=2}} Therefore, our proposal requires the rank two MTC should enhance.
    
    Can one emergent TDL make the IR MTC satisfy two requirements? According to \cite{GK94}, there is only one rank three MTC with $\mbb Z_2$ fusion ring. It has $SU(2)_2$ realization, and has central charge $c=\frac{2n+1}2$ with $n<4$. Can the rank three MTC describe IR CFT? The answer is no.

    A direct way to show this is as follows. From the RG invariance of quantum dimensions, we have $d_j=-1$. Combining this with the fusion ring $jN=N$, we conclude $d_N=0$. This violates invertibility of $F$-symbols.
    
    Therefore, we have ruled out the possibility of one emergent TDL. How about two emergent lines? According to the same paper, there are three rank four MTCs with $\mbb Z_2$ fusion ring.
    
    We can rule out the one with $SU(4)_1$ realization as follows. In the rank four MTC, we have the identification $\eta\to\mcal L_1$. Thus we get $d_1=-1$ from the RG invariance of quantum dimensions. The other fusion rules imply
    \[ d_2=\pm i=-d_3. \]
    However, this contradicts reality of quantum dimensions. Thus, we can rule out the rank four MTC.
    
    Hence, we are left with two rank four MTCs, one with $SU(2)_1\times SU(2)_1$ or $SO(8)_1$ realization, or the other with $SU(2)_3$ realization. Although we cannot completely rule out the first scenario, we can give two arguments which both favor the second scenario.
    
    One argument is about the effective central charge. We can constrain $c^\text{eff}$, and find the first scenario should have $c^\text{eff}=0$. The computation goes as follows. The first scenario has three $\mbb Z_2$ objects, $\mcal L_1,\mcal L_2,\mcal L_3$. The $\mbb Z_2$ fusion rules constrain their spins \cite{CLSWY}:
    \[ s_{1,2,3}\in\frac k4+\frac12\mbb Z. \]
    Thus, no matter which primary has the smallest conformal dimension, we can write $\Delta_\text{smallest}=\frac m4$ with $m\in\mbb Z$. (If the $\mbb Z_2$ primaries all have positive conformal dimensions, then the identity gives $\Delta_\text{smallest}=0$ with $m=0$.) Since the rank four MTC has central charge $c=2n$, the effective central charge is given by
    \[ c^\text{eff}=2n-24\frac m4=2n-6m. \]
    The only possible value compatible with the $c^\text{eff}$-theorem is $c^\text{eff}=0$. However, the empirical fact (\ref{ceffbound}) tells us that nontrivial CFT would have non-zero effective central charge. Thus, this scenario would not be massless.
    
    Another argument employs our hero, free energy. As we explained in the introduction, an MTC with global dimension $D$ contributes $T\ln D$ to free energy. Therefore, at a given temperature, we learn MTCs with smaller global dimensions are energetically favored. Hence, we should ask ``Which of the two MTCs can have smaller global dimension?'' As we saw above, the first scenario has global dimension
    \[ D_\text{first}^2=4\times1=4. \]
    On the other hand, the second scenario has global dimension\footnote{The computation goes as follows. According to \cite{GK94}, the rank four MTC has four objects, identity, $\mbb Z_2$ object $\mcal L_1$, and two non-invertible objects $\mcal L_2,\mcal L_3$ obeying Fibonacci fusion rules. (More precisely, $\mcal L_3$ is a fusion of the $\mbb Z_2$ object and the Fibonacci object $\mcal L_2$, $\mcal L_3=\mcal L_1\mcal L_2$.) The first two objects have quantum dimensions one and minus one, respectively. $d_{\mcal L_1}=-1$ is a result of RG invariance of quantum dimensions. The last two objects have quantum dimensions $d_{\mcal L_2}=\frac{1\pm\sqrt5}2=-d_{\mcal L_3}$. Thus, the global dimension is given by
    \[ D_\text{second}^2=1+1+2\times\left(\frac{1\pm\sqrt5}2\right)^2=5\pm\sqrt5. \]}
    \[ D_\text{second}^2=5\pm\sqrt5. \]
    Comparing the two, we find the second scenario with $D^2=5-\sqrt5$ has the smallest free energy. Indeed, the known IR theory $M(3,5)$ is described by the second scenario --- the rank four MTC with $SU(2)_3$ realization --- with this smallest global dimension. The MTC has central charge $c=\frac{2n+1}5$ with $n<10,n\neq2,7$, and the IR theory is realized by $n=-2$, or $c=-\frac35$ and $c^\text{eff}=\frac35$. One can check spin constraints are satisfied by the matching:
    \begin{equation} 
    \begin{array}{ccc}
    \text{UV}:&1&\mcal L_{1,2}=\eta\\
    &\downarrow&\downarrow\\
    \text{IR}:&1&\mcal L_{1,2}=\eta
    \end{array}.\label{M37toM35}
    \end{equation}
    Note that the Fibonacci line preserved in the previous massless flow is emergent, consistent with the reality condition \cite{K22}.
    
    \item $M(4,7)+\phi_{1,2}$\\
    The UV theory has $c_\text{UV}=-\frac{13}{14},c^\text{eff}_\text{UV}=\frac{11}{14}$. The $\phi_{1,2}$-deformation preserves three TDLs $\{1,\mcal L_{5,1},\mcal L_{3,1}\}$. The non-invertible lines $\mcal L_{5,1},\mcal L_{3,1}$ are associated to primaries with conformal dimensions $\frac{10}7,\frac17$, respectively. The double braidings are given by
    \[ \begin{pmatrix}id_1&id_{5,1}&id_{3,1}\\id_{5,1}&e^{\frac{2\pi i}7}id_1\oplus e^{\frac{4\pi i}7}id_{3,1}&e^{-\frac{2\pi i}7}id_{5,1}\oplus e^{-\frac{6\pi i}7}id_{3,1}\\id_{3,1}&e^{-\frac{2\pi i}7}id_{5,1}\oplus e^{-\frac{6\pi i}7}id_{3,1}&e^{-\frac{4\pi i}7}id_1\oplus e^{\frac{2\pi i}7}id_{5,1}\oplus e^{-\frac{2\pi i}7}id_{3,1}\end{pmatrix}. \]
    Taking the quantum trace, we obtain the unnormalized topological $S$-matrix
    \[ \widetilde S_\text{top}=\begin{pmatrix}1&2\sin\frac\pi{14}&-\frac1{2\sin\frac{3\pi}{14}}\\2\sin\frac\pi{14}&2\cos\frac\pi7-1&1\\-\frac1{2\sin\frac{3\pi}{14}}&1&-\frac{\sin\frac\pi7}{\sin\frac{3\pi}7}\end{pmatrix}. \]
    Dividing with quantum dimensions, we get the monodromy charge matrix
    \[ M=\begin{pmatrix}1&1&1\\1&\frac{2\cos\frac\pi7-1}{4\sin^2\frac\pi{14}}&-\frac{\sin\frac{3\pi}{14}}{\sin\frac\pi{14}}\\1&-\frac{\sin\frac{3\pi}{14}}{\sin\frac\pi{14}}&-8\sin\frac\pi{14}\sin^2\frac{3\pi}{14}\end{pmatrix}. \]
    One sees the symmetric centralizer is trivial. Thus the surviving rank three BFC $\mcal C$ is actually an MTC. According to \cite{GK94}, the MTC has $SU(2)_5/\mbb Z_2$ realization with identifications
    \[ 1^\text{here}=1^\text{there},\quad\phi_{5,1}^\text{here}=\phi_1^\text{there},\quad\phi_{3,1}^\text{here}=\phi_2^\text{there}. \]
    Can the rank three MTC $\mcal C$ describe IR CFT? No. We find the rank three MTC cannot be consistent with the $c^\text{eff}$-theorem. To show this, let us predict IR conformal dimensions.
    
    We denote the IR lines as $\mcal L_{5,1}\to j,\mcal L_{3,1}\to k$. We start from the double braiding of $k$ with itself. The $k$-channel requires
    \[ \theta_k^{-1}=e^{-2\pi ih_k^\text{IR}}=e^{\frac{2\pi i}7}, \]
    or
    \[ h_k^\text{IR}=-\frac17\quad(\text{mod }1). \]
    Next, we look at the $j$-channel of the same double braiding. It requires
    \[ \theta_j=e^{2\pi ih_j^\text{IR}}=e^{-\frac{(2+4)\pi i}7}, \]
    or
    \[ h_j^\text{IR}=-\frac37\quad(\text{mod }1). \]
    The ``monotonicity'' imposes further constraints:
    \[ h_j^\text{IR}=\frac47,-\frac37,-\frac{10}7,\dots,\quad h_k^\text{IR}=-\frac17,-\frac87,-\frac{15}7,\dots\ . \]
    Notice that the double braiding relation and the ``monotonicity'' together require $h_k^\text{IR}$ to be negative. With these data, we can compute the topological $S$- and $T$-matrices:
    \[ \widetilde S_\text{top}=\begin{pmatrix}1&2\sin\frac\pi{14}&-\frac1{2\sin\frac{3\pi}{14}}\\2\sin\frac\pi{14}&2\cos\frac\pi7-1&1\\-\frac1{2\sin\frac{3\pi}{14}}&1&-\frac{\sin\frac\pi7}{\sin\frac{3\pi}7}\end{pmatrix},\quad T=\begin{pmatrix}1&0&0\\0&e^{-\frac{6\pi i}7}&0\\0&0&e^{-\frac{2\pi i}7}\end{pmatrix}. \]
    Using the matrices, we can compute the central charge via the relation\footnote{Another way to compute the central charge is via the Gauss sum:
    \[ e^{\pi ic/2}=\frac{\Omega^+}{\Omega^-}, \]
    where
    \[ \Omega^\pm:=\sum_jd_j^2\theta_j^\pm. \]
    However, this method gives worse accuracy, mod $4$ and not mod $8$.}
    \[ (S_\text{top}T)^3=e^{\frac{\pi ic}4}(S_\text{top})^2. \]
    The result is $e^{\pi ic/4}=e^{-\pi i/7}$ (for $D>0$) and $e^{\pi ic/4}=e^{6\pi i/7}$ (for $D<0$). Solving the equations for central charges, we get
    \[ c_{D>0}=-\frac47\quad(\text{mod }8), \]
    or
    \[ c_{D<0}=\frac{24}7\quad(\text{mod }8). \]
    Note that these are in accord with \cite{GK94}, $c=\frac{4n'}7$ with $n'<7$. Let us compute the effective central charges. The two signs of $D$ can be treated simultaneously by writing $c=\frac{24}7-4n$ with $n\in\mbb N$. Denoting $h_j^\text{IR}=\frac47-l,h_k^\text{IR}=-\frac17-m$ with $l,m\in\mbb N$, we get the effective central charge (since $h_k^\text{IR}<0$, $0$ cannot be the minimum)
    \[ c^\text{eff}=\left(\frac{24}7-4n\right)-24\min\left(\frac47-l,-\frac17-m\right)=\frac47\Big\{(6-7n)-\min(24-42l,-6-42m)\Big\}. \]
    The $c^\text{eff}$-theorem imposes $0\le c^\text{eff}\le\frac{11}{14}$. From the last expression above, the only possible values are $0,\frac47$. This means the bracket $\{\}$ of $c^\text{eff}$ is either $0$ or $1$. One can show there is no solution. Let us denote the value of the bracket $\{\}$ as $b$, i.e., $b=0,1$. We thus try to find integral solutions to
    \begin{equation}
        b=6-7n-\min(24-42l,-6-42m)\label{ceffM47rank3}
    \end{equation}
    with $b=0,1$. If $\Delta_\text{smallest}=h_j^\text{IR}$, this reduces to
    \[ 18+b=7(-n+6l). \]
    For the values of $b$, there is no solution because the RHS is a multiple of seven, while the LHS is not. Similarly, if $\Delta_\text{smallest}=h_k^\text{IR}$, the equation reduces to
    \[ -12+b=7(-n+6m). \]
    This has no solution, either.
    
    We conclude the rank three MTC $\mcal C$ cannot be consistent with the $c^\text{eff}$-theorem. Therefore, our proposal claims the symmetry should enhance.
    
    How large the IR MTC should be? According to \cite{GK94}, there is no rank four and five MTC containing the surviving BFC $\mcal C$. The first candidate appears at rank six. There are two rank six MTCs enlarging $\mcal C$. They have $SU(2)_5$ realization with identifications
    \[ 1^\text{here}=1^\text{there},\quad\phi_j^\text{here}=\phi_2^\text{there},\quad\phi_k^\text{here}=\phi_3^\text{there}, \]
    or $SU(2)_3/\mbb Z_2\times SU(2)_5/\mbb Z_2$ realization with identifications
    \[ 1^\text{here}=1^\text{there},\quad\phi_j^\text{here}=\phi_2^\text{there},\quad\phi_k^\text{here}=\phi_3^\text{there}. \]
    Which of these MTCs are favored? To see which MTC has the smallest free energy, we first compute global dimensions. The given fusion rules and RG invariance of quantum dimensions yield
    \[ D_\text{first}^2=2\times\frac7{4\cos^2\frac\pi{14}}, \]
    and
    \[ D_\text{second}^2=\frac{5\pm\sqrt5}2\times\frac7{4\cos^2\frac\pi{14}}. \]
    We find the second scenario with $D_\text{second}^2=\frac{5-\sqrt5}2\times\frac7{4\cos^2\frac\pi{14}}$ has the smallest free energy, and the first scenario has the second smallest free energy. Is the minimal choice allowed? A detailed study shows it cannot be consistent with the $c^\text{eff}$-theorem, hence rules out the scenario.
    
    To prove this, first notice that the second scenario means it is a Deligne product. Thus, even though half of the objects are emergent, one can easily compute modular data. Here, in order to realize $D_\text{second}^2=\frac{5-\sqrt5}2\times\frac7{4\cos^2\frac\pi{14}}$, the emergent Fibonacci line should have quantum dimension $d_1=\frac{1-\sqrt5}2$. Using these, one can also easily compute central charges.
    The result depends on the sign of $D_\text{second}=\pm\sqrt{\frac{5-\sqrt5}2}$ and the topological twist $\theta$ of the Fibonacci line. The only allowed values of $\theta$ are given by
    \[ \theta=e^{\pm\frac{2\pi i}5}. \]
    We summarize the resulting central charges in the table \ref{c472nd}:
    \begin{table}[H]
    \begin{center}
    \begin{tabular}{c|c|c}
    $D_\text{second}\backslash\theta$&$e^{+\frac{2\pi i}5}$&$e^{-\frac{2\pi i}5}$\\\hline
    $+\sqrt{\frac{5-\sqrt5}2}$&$-\frac6{35}$&$-\frac{34}{35}$\\
    $-\sqrt{\frac{5-\sqrt5}2}$&$\frac{134}{35}$&$\frac{106}{35}$
    \end{tabular}.
    \end{center}
    \caption{Central charges (mod $8$) of the rank six MTC}\label{c472nd}
    \end{table}
    As expected, these are in accord with \cite{GK94}, $c=\frac{4n'+2}{35}$ with $n'<35$ and some $n'$ excluded. From the upper bound we can write $c_+=\frac{134}{35}-4n$ (for $\theta=e^{+\frac{2\pi i}5}$) and $c_-=\frac{106}{35}-4n$ (for $\theta=e^{-\frac{2\pi i}5}$) for $n\in\mbb N$. (These correspond to allowed $n'$'s.)
    
    We are now ready to compute effective central charges. We start from the case $\theta=e^{+\frac{2\pi i}5}$. In this case, in addition to $h_j^\text{IR}=\frac47-l,h_k^\text{IR}=-\frac17-m$ with $l,m\in\mbb N$, we have three emergent lines\footnote{In this scenario, the three emergent lines are the Fibonacci line $\mcal L_1$, $\mcal L_4=j\mcal L_1$, and $\mcal L_5=k\mcal L_1$.} corresponding to primaries with conformal dimensions $h_1=\frac15+p,h_4=-\frac8{35}+q,h_5=\frac2{35}+r$ with $p,q,r\in\mbb Z$. Thus, the effective central charge is given by
    \begin{align*}
        c_+^\text{eff}&=\left(\frac{134}{35}-4n\right)-24\min\left(\frac47-l,-\frac17-m,\frac15+p,-\frac8{35}+q,\frac2{35}+r\right)\\
        &=\frac2{35}\Big\{67-70n-12\min(20-35l,-5-35m,7+35p,-8+35q,2+35r)\Big\}.
    \end{align*}
    Since the bracket $\{\}$ is an (odd) integer, the $c^\text{eff}$-theorem $0\le c^\text{eff}\le\frac{11}{14}$ allows only $\{\}=1,3,\dots,13$. Let us denote the value of $\{\}$ as $b$, i.e., $b=1,3,\dots,13$. Then we try to find integral solutions to
    \begin{equation}
        b=67-70n-12\min(20-35l,-5-35m,7+35p,-8+35q,2+35r).\label{M47rank6+}
    \end{equation}
    We find there is no solution. To show this, we perform case analysis. We start from $\Delta_\text{smallest}=h_j^\text{IR}$. In this case, the equation reduces to
    \[ 173+b=70(-n+6l). \]
    For the values of $b$, the LHS cannot be a multiple of $70$, thus there is no solution. The other cases can be studied in the same way. The case $\Delta_\text{smallest}=h_k^\text{IR}$ gives
    \[ -127+b=70(-n+6m), \]
    the case $\Delta_\text{smallest}=h_1$ gives
    \[ 17+b=70(-n-6p), \]
    the case $\Delta_\text{smallest}=h_4$ gives
    \[ -163+b=70(-n-6q), \]
    and the case $\Delta_\text{smallest}=h_5$ gives
    \[ -43+b=70(-n-6r). \]
    None of them has a solution.
    
    The minus sign $\theta=e^{-\frac{2\pi i}5}$ can be studied in the same manner. Writing $h_1=-\frac15+p,h_4=\frac{13}{35}+q,h_5=-\frac{12}{35}+r$ with $p,q,r\in\mbb Z$, we get the effective central charge
    \begin{align*}
        c^\text{eff}_-&=\left(\frac{106}{35}-4n\right)-24\min\left(\frac47-l,-\frac17-m,-\frac15+p,\frac{13}{35}+q,-\frac{12}{35}+r\right)\\
        &=\frac2{35}\Big\{53-70n-12\min(20-35l,-5-35m,-7+35p,13+35q,-12+35r)\Big\}.
    \end{align*}
    The $c^\text{eff}$-theorem only allows $b:=\{\}=1,3,\dots,13$. We thus try to solve
    \begin{equation}
        b=53-70n-12\min(20-35l,-5-35m,-7+35p,13+35q,-12+35r)\label{M47rank6-}
    \end{equation}
    with these values of $b$. We find there is no solution. To prove this, we again perform case analysis. The case $\Delta_\text{smallest}=h_j^\text{IR}$ gives
    \[ 187+b=70(-n+6l). \]
    This equation does not have solution for the values of $b$. Similarly, the case $\Delta_\text{smallest}=h_k^\text{IR}$ gives
    \[ -113+b=70(-n+6m), \]
    the case $\Delta_\text{smallest}=h_1$ gives
    \[ -137+b=70(-n-6p), \]
    the case $\Delta_\text{smallest}=h_4$ gives
    \[ 103+b=70(-n-6q), \]
    and the case $\Delta_\text{smallest}=h_5$ gives
    \[ -197+b=70(-n-6r). \]
    None of these equations has a solution.
    
    Thus we conclude the second scenario --- the rank six MTC with $SU(2)_2/\mbb Z_2\times SU(2)_5/\mbb Z_2$ realization having global dimension $D_\text{second}^2=\frac{5-\sqrt5}2\times\frac7{4\cos^2\frac\pi{14}}$ --- is inconsistent with the $c^\text{eff}$-theorem.
    
    The next smallest free energy is realized by the first scenario with $SU(2)_5$ realization. Therefore, the rank six MTC is likely to describe the IR CFT, and indeed the known IR theory $M(3,7)$ is described by the MTC. (This fact in particular means the MTC can be consistent with the $c^\text{eff}$-theorem with $c^\text{eff}_\text{IR}=\frac57<\frac{11}{14}$.) One can check spin contents are beautifully matched by the identifications:
    \begin{equation} 
    \begin{array}{cccc}
    \text{UV}:&1&\mcal L_{3,1}&\mcal L_{5,1}\\
    &\downarrow&\downarrow&\downarrow\\
    \text{IR}:&1&\mcal L_{3,1}&\mcal L_{5,1}
    \end{array}.\label{M47toM37}
    \end{equation}
    \item
    
    $M(4,9)+\phi_{5,1}$\\
    The UV theory has $c_\text{UV}=-\frac{19}6,c_\text{UV}^\text{eff}=\frac56$. The $\phi_{5,1}$-deformation preserves three TDLs $\{1,\mcal L_{1,2},\mcal L_{1,3}=\eta\}$. The $\mbb Z_2$ line $\eta$ is associated to a primary with $h=\frac72$, and the non-invertible line $\mcal L_{1,2}$ to a primary with $h=\frac{19}{16}$. The double braidings are given by
    \[ \begin{pmatrix}id_1&id_{1,2}&id_\eta\\id_{1,2}&e^{-\frac{3\pi i}4}id_1\oplus e^{\frac{\pi i}4}id_\eta&-id_{1,2}\\id_\eta&-id_{1,2}&id_1\end{pmatrix}. \]
    Taking the quantum trace, we obtain the unnormalized topological $S$-matrix
    \[ \widetilde S_\text{top}=\begin{pmatrix}1&-\sqrt2&1\\-\sqrt2&0&\sqrt2\\1&\sqrt2&1\end{pmatrix}. \]
    Dividing with the quantum dimensions, we get the monodromy charge matrix
    \[ M=\begin{pmatrix}1&1&1\\1&0&-1\\1&-1&1\end{pmatrix}. \]
    One sees there is no nontrivial transparent line. Hence, the surviving rank three BFC is actually an MTC. According to \cite{GK94}, the MTC has $SU(2)_2$ realization with identifications
    \[ 1^\text{here}=1^\text{there},\quad\phi_{1,2}^\text{here}=\phi_2^\text{there},\quad\phi_\eta^\text{here}=\phi_1^\text{there}. \]
    Can the rank three MTC describe IR CFT? No. We find the rank three MTC cannot be consistent with the $c^\text{eff}$-theorem. To show this, we follow the same procedure as in the last example. Namely, we compute possible effective central charge by predicting IR conformal dimensions. With notations $\mcal L_{1,2}\to j,\eta\to k$, the double braiding of $\mcal L_{1,2}$ with itself predicts
    \[ c_{j,j}^\text{IR}c_{j,j}^\text{IR}=e^{\frac{3\pi i}4}id_1\oplus e^{-\frac{\pi i}4}id_k. \]
    The $k$-channel tells us
    \[ \theta_k=e^{2\pi ih_k^\text{IR}}=e^{-\frac{(1+3)\pi i}4}, \]
    or
    \[ h_k^\text{IR}=\frac12\quad(\text{mod }1). \]
    Similarly, the identity-channel gives
    \[ \theta_j^{-2}=e^{-4\pi ih_j^\text{IR}}=e^{\frac{3\pi i}4}, \]
    or
    \[ h_j^\text{IR}=-\frac3{16}\quad(\text{mod }\frac12). \]
    Combined with the ``monotonicity,'' we get
    \[ h_j^\text{IR}=\frac{13}{16},\frac5{16},-\frac3{16},\dots,\quad h_k^\text{IR}=\frac72,\frac52,\frac32,\dots\ . \]
    With these data, we can compute central charge of the MTC. For $h_j^\text{IR}=\frac{13}{16}$ mod $1$, the topological $S$- and $T$-matrices are given by
    \[ S_\text{top}=\pm\frac12\begin{pmatrix}1&-\sqrt2&1\\-\sqrt2&0&\sqrt2\\1&\sqrt2&1\end{pmatrix},\quad T_1=\begin{pmatrix}1&0&0\\0&e^{-\frac{3\pi i}8}&0\\0&0&-1\end{pmatrix}. \]
    Thus the MTC has central charge $e^{\pi ic/4}=\pm e^{-3\pi i/8}$, or
    \[ c_1=-\frac32\quad(\text{mod }8) \]
    for $D>0$ and
    \[ c_1=\frac52\quad(\text{mod }8) \]
    for $D<0$. For another class $h_j^\text{IR}=\frac5{16}$ mod $1$, the $T$-matrix becomes
    \[ T_2=\begin{pmatrix}1&0&0\\0&e^{\frac{5\pi i}8}&0\\0&0&-1\end{pmatrix}. \]
    Thus the MTC has central charge $e^{\pi ic/4}=\pm e^{5\pi i/8}$, or
    \[ c_2=\frac52\quad(\text{mod }8) \]
    for $D>0$ and
    \[c_2=-\frac32\quad(\text{mod }8) \]
    for $D<0$. These are in accord with \cite{GK94}, $c=\frac{2n'+1}2$ with $n'<4$. From the upper bound, we can write $c=\frac52-4n$ with $n\in\mbb N$ for all cases. Let us see whether these data can satisfy the $c^\text{eff}$-theorem. We start from the case with $T_1$. Denoting $h_j^\text{IR}=\frac{13}{16}-l,h_k^\text{IR}=\frac72-m$ with $l,m\in\mbb N$, we get the effective central charge ($\Delta_\text{smallest}=0$ cannot be consistent with the $c^\text{eff}$-theorem)
    \[ c^\text{eff}=\left(\frac52-4n\right)-24\min\left(\frac{13}{16}-l,\frac72-m\right)=\frac12\Big\{(5-8n)-3\min(13-16l,56-16m)\Big\}. \]
    The $c^\text{eff}$-theorem imposes $0\le c^\text{eff}\le\frac56$. From the expression above, the only possible values are $0,\frac12$. Equivalently, the allowed values of the bracket $\{\}$ is $b=0,1$. We try to find integral solutions to
    \begin{equation}
        b=(5-8n)-3\min(13-16l,56-16m).\label{M49rank3T1}
    \end{equation}
    If $\Delta_\text{smallest}=h_j^\text{IR}$, the equation reduces to
    \[ 34+b=8(-n+6l). \]
    This equation has no solution for the values of $b$. Similarly, if $\Delta_\text{smallest}=h_k^\text{IR}$, (\ref{M49rank3T1}) reduces to
    \[ 163+b=8(-n+6m). \]
    This equation does not have solution, either.
    
    The case with $T_2$ can be studied in the same manner. Denoting $h_j^\text{IR}=\frac5{16}-l$ with $l\in\mbb N$, we get the effective central charge
    \[ c^\text{eff}=\left(\frac52-4n\right)-24\min\left(\frac5{16}-l,\frac72-m\right)=\frac12\Big\{(5-8n)-3\min(5-16l,56-16m)\Big\}. \]
    We thus try to find integral solutions to
    \begin{equation}
        b=(5-8n)-3\min(5-16l,56-16m).\label{M49rank3T2}
    \end{equation}
    If $\Delta_\text{smallest}=h_j^\text{IR}$, this reduces to
    \[ 10+b=8(-n+6l). \]
    This has no solution. The case $\Delta_\text{smallest}=h_k^\text{IR}$ is the same as in the previous case, and there is no solution.
    
    We conclude the rank three MTC cannot be consistent with the $c^\text{eff}$-theorem. Therefore, our proposal requires symmetry enhancement.
    
    How large the IR MTC should be? According to \cite{GK94}, there is no rank four and five MTC containing the surviving TY fusion category. The first candidate appears at rank six. There is only one rank six MTC containing the TY fusion category. It has $SU(2)_3/\mbb Z_2\times SU(2)_2$ realization with identifications
    \[ 1^\text{here}=1^\text{there},\quad\phi_{1,2}^\text{here}=\phi_3^\text{there},\quad\phi_\eta^\text{here}=\phi_2^\text{there}. \]
    Since the MTC is a Deligne product, we can compute effective central charge as in the last example. In the end, we find $c^\text{eff}=\frac15$ is the only possibility. The computation is straightforward, but long. Hence, we relegate the details to the Appendix \ref{M49rank6}.
    
    Can the MTC describe IR CFT? To the best of our knowledge, the classification of non-unitary RCFTs is incomplete, and we do not know if there exists a non-unitary\footnote{The theory, if existed, cannot be unitary. This is because, in that case $\Delta_\text{smallest}=0$, and the smallest central charge in unitary CFT is $\frac12$ realized by the critical Ising model. In other words, in unitary CFTs, there is a lower bound $\frac12\le c^\text{eff}=c$.} RCFT with $c^\text{eff}=\frac15$. What we notice is that the value is smaller than the empirical lower bound $\frac25$ realized by the Lee-Yang model. Therefore, we believe such a non-unitary RCFT does not exist, and we also discard the scenario of rank six MTC. Thus, we keep searching for consistent MTCs with larger ranks. Unfortunately, however, higher rank MTCs are poorly classified, and we could not find\footnote{There is a partial classification up to $r=9$ \cite{W15}. We checked none of the rank $r=7,8,9$ in the list can be the IR MTC because they cannot match the surviving TDL with quantum dimension $\sqrt2$ (more precisely $-\sqrt2$).} a suitable MTC describing the IR RCFT. Any way, the very facts that the surviving rank three MTC has to be enlarged, and the IR MTC should have rank $r>6$ are consistent with the known RG flow to $M(4,7)$. One can see spin contents are matched by the identifications:
    \begin{equation} 
    \begin{array}{cccc}
    \text{UV}:&1&\mcal L_{1,2}&\mcal L_{1,3}=\eta\\
    &\downarrow&\downarrow&\downarrow\\
    \text{IR}:&1&\mcal L_{1,2}&\mcal L_{1,3}=\eta
    \end{array}.\label{M49toM47}
    \end{equation}
    Note in particular that non-invertible TDLs preserved in the previous example are emergent, consistent with the reality condition.
\end{itemize}

\appendix
\setcounter{section}{0}
\renewcommand{\thesection}{\Alph{section}}
\setcounter{equation}{0}
\renewcommand{\theequation}{\Alph{section}.\arabic{equation}}

\section{Lemma, modularity, matching, and ``monotonicity''}\label{statements}
In this appendix, we prove five statements. We start from the following lemma:\newline

\textbf{Lemma.} \textit{The $\phi_{1,2}$-deformation of $M(p,2p-1)$ minimal model preserves $(p-1)$ TDLs $\{\mcal L_{1,1},\mcal L_{3,1},\dots,\mcal L_{2p-3,1}\}$. The $\phi_{5,1}$-deformation of $M(p,2p+1)$ minimal model preserves $(p-1)$ TDLs $\{\mcal L_{1,1},\mcal L_{1,2},\dots,\mcal L_{1,p-1}\}$.}

\textit{Proof.}

In $M(p,2p-1)$ theory, the action of $\mcal L_{r,s}$ on $\phi_{1,2}$ is given by
\[ \hat{\mcal L}_{r,s}|\phi_{1,2}\ra=\frac{S_{rs,12}}{S_{11,12}}|\phi_{1,2}\ra=(-1)^{1+s}\frac{\sin\frac{\pi pr}{2p-1}\sin\frac{2\pi s}p}{\sin\frac{\pi p}{2p-1}\sin\frac{2\pi} p}|\phi_{1,2}\ra. \]
Thus, Kac indices of surviving TDLs can be written as
\begin{equation}
    \widetilde E_{p,2p-1}^{\phi_{1,2}}:=\left\{(r,s)\in E_{p,2p-1}\Big|\frac{\sin\frac{\pi s}p}{\sin\frac\pi p}=(-1)^{1+r}\frac{\sin\frac{2\pi s}p}{\sin\frac{2\pi}p}\right\}.\label{E12}
\end{equation}
Employing the double-angle formula, the defining condition reduces to
\begin{equation}
    \cos\frac\pi p=(-1)^{1+r}\cos\frac{\pi s}p=\cos\pi\left(1+r+\frac sp\right).\label{E12'}
\end{equation}
This can be easily solved:
\[ \pi\left(1+r+\frac sp\right)=\pm\frac\pi p+2\pi n\quad (n\in\mbb Z). \]
When $r$ is odd, this is equivalent to
\[ \frac sp=\pm\frac1p+2n, \]
and the only solution is $s=1$ with plus sign. When $r$ is even, the equation is equivalent to
\[ \frac sp=1\pm\frac1p+2n, \]
and the only solution is $s=p-1$ with minus sign. This proves
\[ \widetilde E_{p,2p-1}^{\phi_{1,2}}=\left\{(r,s)\in E_{p,2p-1}\Big|(2R+1,1),(2R+2,p-1)\right\}. \]
Here, recall the equivalence relation $(r,s)\sim(2p-1-r,p-s)$. Then we notice the two solutions are equivalent:
\[ (2R+2,p-1)\sim(2(p-R-2)+1,1). \]
In summary, after removing double counting, we arrive the Kac indices of $\phi_{1,2}$-surviving TDLs:
\begin{equation}
    \widetilde E_{p,2p-1}^{\phi_{1,2}}=\left\{(2R+1,1)\Big|R=0,1,\dots,p-2\right\}.\label{E12''}
\end{equation}

In $M(p,2p+1)$ theory, the action of $\mcal L_{r,s}$ on $\phi_{5,1}$ is given by
\[ \hat{\mcal L}_{r,s}|\phi_{5,1}\ra=\frac{S_{rs,51}}{S_{11,51}}|\phi_{5,1}\ra=(-1)^{r+5s}\frac{\sin\frac{5\pi pr}{2p+1}\sin\frac{\pi s}p}{\sin\frac{5\pi p}{2p+1}\sin\frac\pi p}|\phi_{5,1}\ra. \]
Thus, Kac indices of surviving TDLs can be written as
\begin{equation}
    \widetilde E_{p,2p+1}^{\phi_{5,1}}:=\left\{(r,s)\in E_{p,2p+1}\Big|\frac{\sin\frac{\pi pr}{2p+1}}{\sin\frac{\pi p}{2p+1}}=\frac{\sin\frac{5\pi pr}{2p+1}}{\sin\frac{5\pi p}{2p+1}}\right\}.\label{E51}
\end{equation}
Employing the quintuple-angle formula
\[ \sin5a=5\cos^4a\sin a-10\cos^2a\sin^3a+\sin^5a \]
and the double-angle formula, the RHS of the defining condition reduces to
\begin{align*}
    (\text{RHS})&=\frac{\sin r\theta}{\sin\theta}\frac{5\cos^4r\theta-10\cos^2r\theta\sin^2r\theta+\sin^4r\theta}{5\cos^4\theta-10\cos^2\theta\sin^2\theta+\sin^4\theta}\\
    &=\frac{\sin r\theta}{\sin\theta}\frac{1+2\cos2r\theta+2\cos4r\theta}{1+2\cos2\theta+2\cos4\theta}
\end{align*}
with $\theta:=\frac{\pi p}{2p+1}$. Equivalently, the defining condition reduces to
\[ 0=\frac{\sin r\theta}{\sin\theta}\left(\frac{1+2\cos2r\theta+2\cos4r\theta}{1+2\cos2\theta+2\cos4\theta}-1\right). \]
Since the overall coefficient is non-zero for $r=1,2,\dots,2p$, the condition further reduces to
\[ \cos2r\theta+\cos4r\theta=\cos2\theta+\cos4\theta. \]
We apply the sum-to-product formula to get
\begin{equation}
    0=\sin(r+1)\theta\sin(r-1)\theta\Big[4\cos(r+1)\theta\cos(r-1)\theta+1\Big].\label{51reduced}
\end{equation}
The equation is solved if the overall coefficient is zero or the bracket $[\ ]$ is zero.

The overall coefficient is zero iff
\[ (r\pm1)\theta=n\pi\quad(n\in\mbb Z). \]
This is equivalent to
\[ r\pm1=\left(2+\frac1p\right)n. \]
For $(r\pm1)\in\mbb N$, we need $n=pN$ with $N\in\mbb N$:
\[ r\pm1=(2p+1)N. \]
Since $r\pm1=0,1,\dots,2p+1$, we conclude the only possibility is $N=0,1$ and
\[ r=1,2p. \]
Otherwise, the overall coefficient is non-zero, and (\ref{51reduced}) further simplifies to
\[ \cos(r+1)\theta\cos(r-1)\theta=-\frac14. \]
For the product of two cosines to be negative, one should be positive, and the other negative. Furthermore, since $\theta\lessim\frac\pi2$, $(r-1)\theta$ has to be in the first or third quadrant, and $(r+1)\theta$ has to be in the second or fourth quadrant, respectively. Let us see whether the first case have a solution. The condition is
\[ 2n\pi<(r-1)\theta<\left(\frac12+2n\right)\pi<(r+1)\pi<(1+2n)\pi\quad(n\in\mbb N). \]
This is equivalent to
\[ 1+4n+\frac{2n}p<r<2+4n+\frac{2n}p+\frac1{2p}\ \&\ 4n+\frac{2n}p+\frac1{2p}<r<1+4n+\frac{2n}p+\frac1p. \]
Since $p\ge3$, $1>\frac1{2p}$, and the left most expression gives stronger lower bound. Similarly, the right most expression gives stronger upper bound. We conclude
\[ 1+4n+\frac{2n}p<r<1+4n+\frac{2n}p+\frac1p, \]
however, there is no such natural number. This can be easily understood by multiplying by $p$ to the inequality
\[ p+4pn+2n<pr<p+4pn+2n+1. \]
Thus the first case does not have a solution.

The second case has no solution, either. The condition
\[ (2n+1)\pi<(r-1)\theta<\left(2n+\frac32\right)\pi<(r+1)\theta<(2n+2)\pi\quad(n\in\mbb N) \]
is equivalent to
\[ 3+4n+\frac{2n}p+\frac1p<r<4+4n+\frac{2n}p+\frac3{2p}\ \&\ 2+4n+\frac{2n}p+\frac3{2p}<r<3+4n+\frac{2n}p+\frac2p. \]
The stronger inequalities give
\[ 3+4n+\frac{2n}p+\frac1p<r<3+4n+\frac{2n}p+\frac2p. \]
This has no solution.

We conclude the only solutions to the defining condition is $r=1,2p$, or
\[ \widetilde E_{p,2p+1}^{\phi_{5,1}}=\left\{(r,s)\in E_{p,2p+1}\Big|(1,s),(2p,s)\right\}. \]
Here, recall the equivalence relation $(r,s)\sim(2p+1-r,p-s)$. Then, we have
\[ (2p,s)\sim(1,p-s). \]
In summary, after removing double counting, Kac indices of $\phi_{5,1}$-surviving TDLs are given by
\begin{equation}
    \widetilde E_{p,2p+1}^{\phi_{5,1}}=\left\{(1,s)\Big|s=1,2,\dots,p-1\right\}.\label{E51'}
\end{equation}$\square$\newline

With this knowledge, we can prove surviving BFCs are always modular. This is our second statement:\newline

\textbf{Proposition.}
\textit{In the massless flows $M(p,2p+1)\to M(p,2p-1)$ and $M(p,2p-1)\to M(p-1,2p-1)$, surviving BFCs $\mcal C$'s are modular.}

\textit{Proof.}

To study modularity of suviving BFC $\mcal C$, we employ the following theorem \cite{K05,K22}:
\[ i\in\mcal C'\equiv Z_2(\mcal C)\iff\forall j\in\mcal C,\ M_{ij}:=\frac{\left(S_\text{top}\right)_{ij}\left(S_\text{top}\right)_{11}}{\left(S_\text{top}\right)_{1i}\left(S_\text{top}\right)_{1j}}=1. \]
The $M_{ij}$ is called the monodromy charge matrix. Using the formula of the $S$-matrix (\ref{Smatrix}), we get an explicit formula for the monodromy charge matrices. For the first flow, we have
\begin{equation}
    M_{(1,s),(1,s')}^{M(p,2p+1)}=\frac{\sin\frac{\pi ss'}p\sin\frac\pi p}{\sin\frac{\pi s}p\sin\frac{\pi s'}p},\label{M1st}
\end{equation}
and for the second flow, we have
\begin{equation}
    M_{(2R+1,1),(2R'+1,1)}^{M(p,2p-1)}=\frac{\sin\frac{\pi p(2R+1)(2R'+1)}{2p-1}\sin\frac{\pi p}{2p-1}}{\sin\frac{\pi p(2R+1)}{2p-1}\sin\frac{\pi p(2R'+1)}{2p-1}}.\label{M2nd}
\end{equation}

With these formula, let us try to search for transparent lines. We follow the strategy of \cite{K21}. Namely, we constrain possible transparent lines by studying lines in $\mcal C$ one by one.

We start from the first flow. A line $\mcal L_{1,s_*}$ is transparent in the surviving BFC $\mcal C$ iff for any $\mcal L_{1,s}\in\mcal C$, $M_{(1,s),(1,s_*)}^{M(p,2p+1)}=1$. The equation is trivially satisfied for $(1,s)=(1,1)$, and does not impose any constraint on $s_*$. Next, we study $(1,s)=(1,2)$. In this case, the equation reduces to
\[ 1\stackrel!{=}M_{(1,2),(1,s_*)}^{M(p,2p+1)}=\frac{\sin\frac{2\pi s_*}p\sin\frac\pi p}{\sin\frac{2\pi}p\sin\frac{\pi s_*}p}. \]
The double-angle formula gives much simpler expression:
\[ \cos\frac{\pi s_*}p=\cos\frac\pi p. \]
This equation can be easily solved:
\[ \frac{\pi s_*}p=\pm\frac\pi p+2\pi n\quad(n\in\mbb Z), \]
or
\[ s_*=\pm1+2pn. \]
The only solution in the range $\{1,2,\dots,p-1\}$ is $s_*=1$, i.e., the identity line $1=\mcal L_{1,1}$. We conclude the symmetric centralizer is trivial for the first flow.

Let us study the second flow in the same way. A line $\mcal L_{(2R_*+1,1)}$ is transparent in the surviving BFC $\mcal C$ iff for any $\mcal L_{(2R+1,1)}\in\mcal C$, $M_{(2R+1,1),(2R_*+1,1)}^{M(p,2p-1)}=1$. The equation is trivially satisfied for $(2R+1,1)=(1,1)$, and this case does not constrain $R_*$. Next, let us study $(2R+1,1)=(3,1)$. In this case, the equation reduces to
\[ 1\stackrel!=M_{(3,1),(2R_*+1,1)}^{M(p,2p-1)}=\frac{\sin\frac{3\pi p(2R_*+1)}{2p-1}\sin\frac{\pi p}{2p-1}}{\sin\frac{3\pi p}{2p-1}\sin\frac{\pi p(2R_*+1)}{2p-1}}. \]
The triple-angle formula and double-angle formula give much simpler form:
\[ 1=\frac{1+2\cos\frac{2\pi p(2R_*+1)}{2p-1}}{1+2\cos\frac{2\pi p}{2p-1}}, \]
or
\[ \cos\frac{2\pi p(2R_*+1)}{2p-1}=\cos\frac{2\pi p}{2p-1}. \]
Again, this can be easily solved:
\[ \frac{2\pi p(2R_*+1)}{2p-1}=\pm\frac{2\pi p}{2p-1}+2\pi n\quad(n\in\mbb Z), \]
or
\[ 2R_*+1=\pm1+\left(2-\frac1p\right)n. \]
For the RHS to be an integer, we need $n=pN$ with $N\in\mbb Z$. Then, the only solution in the range $2R_*+1\in\{1,3,\dots,2p-3\}$ is $R_*=0$, i.e., the identity line $1=\mcal L_{1,1}$. We conclude the symmetric centralizer is also trivial for the second flow. $\square$\newline

With the concrete knowledge on surviving lines provided by the lemma, it is also not hard to match TDLs in UV and IR. This is our third statement:\newline

\textbf{Proposition.} \textit{In the massless flows $M(p,2p+1)\to M(p,2p-1)$ triggered by the primary $\phi_{5,1}$, and $M(p,2p-1)\to M(p-1,2p-1)$ triggered by the primary $\phi_{1,2}$, TDLs in UV and IR are matched as
\begin{equation} 
\begin{array}{ccccc}
M(p,2p+1):&\mcal L_{1,1}&\mcal L_{1,2}&\cdots&\mcal L_{1,p-1}\\
&\downarrow&\downarrow&\cdots&\downarrow\\
M(p,2p-1):&\mcal L_{1,1}&\mcal L_{1,2}&\cdots&\mcal L_{1,p-1}
\end{array},\label{Mp2p+1matchingTDL}
\end{equation}
and
\begin{equation} 
\begin{array}{ccccc}
M(p,2p-1):&\mcal L_{1,1}&\mcal L_{3,1}&\cdots&\mcal L_{2p-3,1}\\
&\downarrow&\downarrow&\cdots&\downarrow\\
M(p-1,2p-1):&\mcal L_{1,1}&\mcal L_{3,1}&\cdots&\mcal L_{2p-3,1}
\end{array},\label{Mp2p-1matchingTDL}
\end{equation}
respectively.}

\textit{Proof.}

In the first flow, surviving lines have quantum dimensions
\[ d_{(1,s)}^{M(p,2p+1)}=(-1)^{s+1}\frac{\sin\frac{\pi s}p}{\sin\frac\pi p}, \]
while in IR theory $M(p,2p-1)$, a TDL with Kac index $(t,u)$ has quantum dimension
\[ d_{(t,u)}^{M(p,2p-1)}=(-1)^{t+u}\frac{\sin\frac{\pi pt}{2p-1}\sin\frac{\pi u}p}{\sin\frac{\pi p}{2p-1}\sin\frac\pi p}. \]
We find quantum dimensions are matched by $(t,u)=(1,s),(1,p-s)$. Taking into account the special role played by $\mcal L_{1,2}$ under fusion, we conclude $\mcal L_{1,s}^{M(p,2p+1)}\to\mcal L_{1,s}^{M(p,2p-1)}$.

In the second flow, surviving lines have quantum dimensions
\[ d_{(2R+1,1)}^{M(p,2p-1)}=\frac{\sin\frac{\pi p(2R+1)}{2p-1}}{\sin\frac{\pi p}{2p-1}}, \]
while in IR theory $M(p-1,2p-1)$, a TDL with Kac index $(t,u)$ has quantum dimension
\[ d_{(t,u)}^{M(p-1,2p-1)}=(-1)^{t+u}\frac{\sin\frac{\pi(p-1)t}{2p-1}\sin\frac{\pi u}{p-1}}{\sin\frac{\pi(p-1)}{2p-1}\sin\frac\pi{p-1}}. \]
We find the quantum dimensions are matched by $(t,u)=(2R+1,1)$:
\begin{align*}
    d_{(2R+1,1)}^{M(p-1,2p-1)}&=\frac{\sin\frac{\pi(p-1)(2R+1)}{2p-1}}{\sin\frac{\pi(p-1)}{2p-1}}\\
    &=\frac{\sin\frac{\pi(2p-1-p)(2R+1)}{2p-1}}{\sin\frac{\pi(2p-1-p)}{2p-1}}\\
    &=\frac{\sin\frac{\pi p(2R+1)}{2p-1}}{\sin\frac{\pi p}{2p-1}}.
\end{align*}
We conclude $\mcal L_{2R+1,1}^{M(p,2p-1)}\to\mcal L_{2R+1,1}^{M(p-1,2p-1)}$. $\square$\newline

Once we understand which lines are matched, it is easy to prove the ``monotonicity'' of scaling dimensions under the massless flows. This is our fourth statement:\newline

\textbf{Proposition.}
\textit{Under the massless flows $M(p,2p+1)\to M(p,2p-1)\to M(p-1,2p-1)$, scaling dimensions of surviving lines decrease ``monotonically.''}

\textit{Proof.}

We just compute conformal dimensions. In the first flow $M(p,2p+1)\to M(p,2p-1)$, a surviving line $\mcal L_{1,s}$ is associated to a UV primary with conformal dimension $h_{1,s}^{M(p,2p+1)}=\frac{[p-(2p+1)s]^2-(p+1)^2}{4p(2p+1)}$ and an IR primary with conformal dimension $h_{1,s}^{M(p,2p-1)}=\frac{[p-(2p-1)s]^2-(p-1)^2}{4p(2p-1)}$. Their difference is
\[ h_{1,s}^{M(p,2p+1)}-h_{1,s}^{M(p,2p-1)}=\frac{s^2-1}{2p}\ge0. \]
The equality is saturated by the identity line $1=\mcal L_{1,1}$, and we have strict inequalities for nontrivial lines.

In the second flow $M(p,2p-1)\to M(p-1,2p-1)$, a surviving line $\mcal L_{2R+1,1}$ is associated to a UV primary with conformal dimension $h_{2R+1,1}^{M(p,2p-1)}=\frac{[p(2R+1)-(2p-1)]^2-(p-1)^2}{4p(2p-1)}$ and an IR primary with conformal dimension $h_{2R+1,1}^{M(p-1,2p-1)}=\frac{[(p-1)(2R+1)-(2p-1)]^2-p^2}{4(p-1)(2p-1)}$. Their difference is
\[ h_{2R+1,1}^{M(p,2p-1)}-h_{2R+1,1}^{M(p-1,2p-1)}=\frac{R(R+1)}{2p-1}\ge0. \]
The equality is saturated by the identity line $1=\mcal L_{1,1}$, and we have strict inequalities for nontrivial lines. $\square$\newline

In this paper, the topological entanglement entropy $S\ni-\ln D$ (and free energy) was the hero. Since entropy is subject to the second law of thermodynamics, it is natural to expect global dimensions are non-increasing along RG flows. In fact, if we naively compare only topological entanglement entropies in UV and IR, we can prove the expectation. This is our final statement:\newline

\textbf{Proposition.}
\textit{Under the massless flows $M(p,2p+1)\to M(p,2p-1)\to M(p-1,2p-1)$, global dimensions $D^2$ decrease ``monotonically.'' Similarly, under the unitary massless flows $M(m+1,m)\to M(m,m-1)$, global dimensions $D$ decrease ``monotonically.''}

\textit{Proof.}

From the formula of the $S$-matrix, the bosonic minimal model $M(p,q)$ has global dimension\footnote{The non-unitary minimal models have alternative signs for $D$. However, their magnitudes are still ``monotonically'' decreasing. This is why we take square of $D$ in non-unitary theories.}
\[ D_{M(p,q)}^2=\frac{pq}8\frac1{\sin^2\frac{\pi p}q\sin^2\frac{\pi q}p}. \]
Ratios of this give the desired results. We start from the flow $M(p,2p+1)\to M(p,2p-1)$. Their ratio in UV and IR is given by
\[ \frac{D_{M(p,2p+1)}^2}{D_{M(p,2p-1)}^2}=\frac{2p+1}{2p-1}\frac{\sin^2\frac{\pi p}{2p-1}}{\sin^2\frac{\pi p}{2p+1}}. \]
The RHS is larger than one for $p\ge3$, showing the result. Similarly, in the massless flow $M(p,2p-1)\to M(p-1,2p-1)$, the ratio is given by
\[ \frac{D_{M(p,2p-1)}^2}{D_{M(p-1,2p-1)}^2}=\frac p{p-1}\frac{\sin^2\frac\pi{p-1}}{\sin^2\frac\pi p}. \]
The RHS is larger than one for $p\ge3$.

The massless RG flows among unitary minimal models can be studied in the same way. The minimal model $M(m+1,m)$ has global dimension
\[ D_{M(m+1,m)}=\frac1{\sqrt{\frac8{m(m+1)}}\sin\frac\pi m\sin\frac\pi{m+1}}. \]
Thus, under the massless flow $M(m+1,m)\to M(m,m-1)$, their ratio in UV and IR is given by
\[ \frac{D_{M(m+1,m)}}{D_{M(m,m-1)}}=\sqrt{\frac{m+1}{m-1}}\frac{\sin\frac\pi{m-1}}{\sin\frac\pi{m+1}}. \]
The RHS is larger than one for $m\ge4$. This proves the statement for this class of massless RG flows. $\square$

\section{Spin contents}\label{spin}
In this appendix, we list spin contents relevant for our study.

\subsection{$M(2,5)$}
\[ \mcal H_{\mcal L_{3,1}}:\quad s\in\{0,\pm\frac15\}\text{ mod }1. \]

\subsection{$M(3,5)$}
\begin{align*}
    \mcal H_{\mcal L_{1,2}}:&\quad s\in\{\pm\frac14\}\text{ mod }1,\\
    \mcal H_{\mcal L_{3,1}}:&\quad s\in\{0,\pm\frac15\}\text{ mod }1.
\end{align*}

\subsection{$M(3,7)$}
\begin{align*}
    \mcal H_{\mcal L_{1,2}}:&\quad s\in\{\pm\frac14\}\text{ mod }1,\\
    \mcal H_{\mcal L_{3,1}}:&\quad s\in\{0,\pm\frac17,\pm\frac27\}\text{ mod }1,\\
    \mcal H_{\mcal L_{5,1}}:&\quad s\in\{0,\pm\frac27,\pm\frac47\}\text{ mod }1.
\end{align*}

\subsection{$M(4,7)$}
\begin{align*}
    \mcal H_{\mcal L_{1,2}}:&\quad s\in\{\pm\frac3{16},\pm\frac5{16}\}\text{ mod }1,\\
    \mcal H_{\mcal L_{1,3}}:&\quad s\in\{0,\pm\frac12\}\text{ mod }1,\\
    \mcal H_{\mcal L_{3,1}}:&\quad s\in\{0,\pm\frac17,\pm\frac27\}\text{ mod }1,\\
    \mcal H_{\mcal L_{5,1}}:&\quad s\in\{0,\pm\frac27,\pm\frac47\}\text{ mod }1.
\end{align*}

\subsection{$M(4,9)$}
\begin{align*}
    \mcal H_{\mcal L_{1,2}}:&\quad s\in\{\pm\frac3{16},\pm\frac5{16}\}\text{ mod }1,\\
    \mcal H_{\mcal L_{1,3}}:&\quad s\in\{0,\pm\frac12\}\text{ mod }1.
\end{align*}

\section{Calculation of $c^\text{eff}$ in the $SU(2)_3/\mbb Z_2\times SU(2)_2$ scenario}\label{M49rank6}
In this appendix, we show the only consistent value of the effective central charge is $c^\text{eff}=\frac15$ in the $SU(2)_3/\mbb Z_2\times SU(2)_2$ scenario. We prove it via direct computation.
    
We denote the Fibonacci line $W$. The other two emergent lines are given by $kW,jW$. (Recall our notation $\mcal L_1\to j,\eta\to k$.) From fusion rules, they have quantum dimensions
\[ d_W=\frac{1\pm\sqrt5}2=d_{kW},\quad d_{jW}=-\sqrt2d_W. \]
Let us start from the case $d_W=\frac{1+\sqrt5}2$. Its allowed topological twists are
\[ \theta_W=e^{\pm\frac{4\pi i}5}. \]
Depending on the sign $D_\text{Fibonacci}=\pm\sqrt{\frac{5+\sqrt5}2}$, the topological twist, and $T_1$ vs. $T_2$, we get the following central charges:
\begin{table}[H]
\begin{center}
\begin{tabular}{c|c|c}
$D_\text{Fibonacci}\backslash\theta_W$&$e^{+\frac{4\pi i}5}$&$e^{-\frac{4\pi i}5}$\\\hline
$+\sqrt{\frac{5+\sqrt5}2}$&$\frac{13}{10}$&$\frac{37}{10}$\\
$-\sqrt{\frac{5+\sqrt5}2}$&$-\frac{27}{10}$&$-\frac3{10}$
\end{tabular},
\end{center}
\caption{Central charges (mod $8$) of the rank six MTC with $T_1$}\label{c49T1plus}
\end{table}
\hspace{-18pt}and
\begin{table}[H]
\begin{center}
\begin{tabular}{c|c|c}
$D_\text{Fibonacci}\backslash\theta_W$&$e^{+\frac{4\pi i}5}$&$e^{-\frac{4\pi i}5}$\\\hline
$+\sqrt{\frac{5+\sqrt5}2}$&$-\frac{27}{10}$&$-\frac3{10}$\\
$-\sqrt{\frac{5+\sqrt5}2}$&$\frac{13}{10}$&$\frac{37}{10}$
\end{tabular}.
\end{center}
\caption{Central charges (mod $8$) of the rank six MTC with $T_2$}\label{c49T2plus}
\end{table}
\hspace{-18pt}As expected, all values are in accord with \cite{GK94}: $c=\frac{2n'+1}{10}$ with $n'<20$ and $n'\neq2,7,12,17$.
Let us see whether these values give effective central charges consistent with the $c^\text{eff}$-theorem.

From the upper bound, we can write $c=\frac{13}{10}-4n$ (for $\theta_W=e^{+\frac{4\pi i}5}$) and $c=\frac{37}{10}-4n$ (for $\theta_W=e^{-\frac{4\pi i}5}$) with $n\in\mbb N$. More precisely, for $T_1$, $n$ is even (resp. odd) for $D_\text{Fibonacci}>0$ (resp. $D_\text{Fibonacci}<0$), and for $T_2$, $n$ is odd (resp. even) for $D_\text{Fibonacci}>0$ (resp. $D_\text{Fibonacci}<0$). The conformal dimensions corresponding to emergent lines depend on the topological twist and $T_1$ vs. $T_2$. Hence, we perform case analysis.
\newline

\underline{i) $T_1$ and $\theta_W=e^{+\frac{4\pi i}5}$}:\\
In this case, the Fibonacci line corresponds to a primary with $h_1=\frac25+p$ with $p\in\mbb Z$. Accordingly, the other emergent lines $kW,jW$ correspond to $h_4=-\frac1{10}+q,h_5=\frac{17}{80}+r$ with $q,r\in\mbb Z$, respectively. Thus, the effective central charge is given by ($\Delta_\text{smallest}=0$ cannot give effective central charge consistent with the $c^\text{eff}$-theorem)
\begin{align*}
    c^\text{eff}&=\left(\frac{13}{10}-4n\right)-24\min\left(\frac{13}{16}-l,\frac72-m,\frac25+p,-\frac1{10}+q,\frac{17}{80}+r\right)\\
    &=\frac1{10}\Big\{(13-40n)-3\min(65-80l,280-80m,32+80p,-8+80q,17+80r)\Big\}.
\end{align*}
The $c^\text{eff}$-theorem imposes $0\le c^\text{eff}\le\frac56$. From the expression above, the bracket $\{\}$ is only allowed to take $b=0,1,\dots,8$. Thus, we try to solve
\begin{equation}
    b=(13-40n)-3\min(65-80l,280-80m,32+80p,-8+80q,17+80r)\label{M49rank61}
\end{equation}
with these values of $b$. If $\Delta_\text{smallest}=h_j^\text{IR}$, the equation reduces to
\[ 182+b=40(-n+6l). \]
The RHS is a multiple of $40$, while the LHS is not for the allowed values of $b$. The equation does not have a solution. The other cases can be studied in the same way. The case $\Delta_\text{smallest}=h_k^\text{IR}$ gives
\[ 827+b=40(-n+6m), \]
the case $\Delta_\text{smallest}=h_1$ gives
\[ 83+b=40(-n-6p), \]
the case $\Delta_\text{smallest}=h_4$ gives
\[ -37+b=40(-n-6q), \]
and the case $\Delta_\text{smallest}=h_5$ gives
\[ 38+b=40(-n-6r). \]
All but the last equations do not have a solution. The last equation seems to have a solution at $b=2$. Let us thus study this case in more detail. For $D_\text{Fibonacci}>0$, $n$ is even, and the RHS becomes $80(-\frac n2-3r)$. There is no solution. For $D_\text{Fibonacci}<0$, $n$ is odd, and with the notation $n=2N+1$ with $N\in\mbb N$, the equation reduces to
\[ 78+b=80(-N-3r). \]
This equation does have solutions at $b=2$ and $N=-3r-1$, or $n=-6r-1=5,11,17,\dots$ for $r=-1,-2,-3,\dots$ . (Since $h_5<0$, it is consistent with $\Delta_\text{smallest}<0$.) The value $b=2$ means $c^\text{eff}=\frac15$.

\underline{ii) $T_1$ and $\theta_W=e^{-\frac{4\pi i}5}$}:\\
The other cases can be studied in exactly the same way. In this case, conformal dimensions become $h_1=-\frac25+p,h_4=\frac1{10}+q,h_5=\frac{33}{80}+r$. Thus the effective central charge is given by ($\Delta_\text{smallest}=0$ cannot be consistent with the $c^\text{eff}$-theorem)
\begin{align*}
    c^\text{eff}&=\left(\frac{37}{10}-4n\right)-24\min\left(\frac{13}{16}-l,\frac72-m,-\frac25+p,\frac1{10}+q,\frac{33}{80}+r\right)\\
    &=\frac1{10}\Big\{(37-40n)-3\min(65-80l,280-80m,-32+80p,8+80q,33+80r)\Big\}.
\end{align*}
Each case reduces to
\begin{align*}
    158+b&=40(-n+6l),\\
    803+b&=40(-n+6m),\\
    -133+b&=40(-n-6p),\\
    -13+b&=40(-n-6q),\\
    62+b&=40(-n-6r).
\end{align*}
All but the first have no solution. Let us look at the first case in detail. For $D_\text{Fibonacci}<0$, $n$ is odd. With the notation $n=2N+1$ with $N\in\mbb N$, the equation reduces to
\[ 198+b=80(-N+3l). \]
This has no solution. For $D_\text{Fibonacci}>0$, $n$ is even, and the RHS becomes $80(-\frac n2+3l)$. This does have a solution at $b=2$ and $n=6l-4=2,8,14,\dots$ . (Since $h_j^\text{IR}<0$, this is consistent with $\Delta_\text{smallest}<0$.) The value $b=2$ gives $c^\text{eff}=\frac15$.

\underline{iii) $T_2$ and $\theta_W=e^{+\frac{4\pi i}5}$}:\\
In this case, the conformal dimensions become $h_1=\frac25+p,h_4=-\frac1{10}+q,h_5=-\frac{23}{80}+r$. Thus, the effective central charge is given by ($\Delta_\text{smallest}=0$ cannot be consistent with the $c^\text{eff}$-theorem)
\begin{align*}
    c^\text{eff}&=\left(\frac{13}{10}-4n\right)-24\min\left(\frac5{16}-l,\frac72-m,\frac25+p,-\frac1{10}+q,-\frac{23}{80}+r\right)\\
    &=\frac1{10}\Big\{(13-40n)-3\min(25-80l,280-80m,32+80p,-8+80q,-23+80r)\Big\}.
\end{align*}
Each case reduces to
\begin{align*}
    62+b&=40(-n+6l),\\
    827+b&=40(-n+6m),\\
    83+b&=40(-n-6p),\\
    -37+b&=40(-n-6q),\\
    -82+b&=40(-n-6r).
\end{align*}
All but the last have no solution. Let us study the last case. For $D_\text{Fibonacci}>0$, $n$ is odd. Writing $n=2N+1$ with $N\in\mbb N$, we get
\[ -42+b=80(-N-3r). \]
There is no solution. For $D_\text{Fibonacci}<0$, $n$ is even, and the RHS reduces to $80(-\frac n2-3r)$. This equation does have solutions at $b=2$ and $n=-6r+2=2,8,14,\dots$ . (Since $h_5<0$, this is consistent with $\Delta_\text{smallest}<0$.) The value $b=2$ means $c^\text{eff}=\frac15$.

\underline{iv) $T_2$ and $\theta_W=e^{-\frac{4\pi i}5}$}:\\
In this case, the conformal dimensions become $h_1=-\frac25+p,h_4=\frac1{10}+q,h_5=-\frac7{80}+r$. Thus, the effective central charge is given by ($\Delta_\text{smallest}=0$ cannot be consistent with the $c^\text{eff}$-theorem)
\begin{align*}
    c^\text{eff}&=\left(\frac{37}{10}-4n\right)-24\min\left(\frac5{16}-l,\frac72-m,-\frac25+p,\frac1{10}+q,-\frac7{80}+r\right)\\
    &=\frac1{10}\Big\{(37-40n)-3\min(25-80l,280-80m,-32+80p,8+80q,-7+80r)\Big\}.
\end{align*}
Each case reduces to
\begin{align*}
    38+b&=40(-n+6l),\\
    803+b&=40(-n+6m),\\
    -133+b&=40(-n-6p),\\
    -13+b&=40(-n-6q),\\
    -58+b&=40(-n-6r).
\end{align*}
All but the first have no solution. Let us see the first case in more detail. For $D_\text{Fibonacci}<0$, $n$ is even, and the RHS reduces to $80(-\frac n2+3l)$. There is no solution. For $D_\text{Fibonacci}>0$, $n$ is odd, $n=2N+1$ with $N\in\mbb N$. Then the equation reduces to
\[ 78+b=80(-N+3l). \] This does have solutions at $b=2$ and $N=3l-1$, or $n=6l-1=5,11,17,\dots$ . The value $b=2$ gives $c^\text{eff}=\frac15$.

This completes the case analysis. In all cases, for $d_W=\frac{1+\sqrt5}2$, we found the only possibility is $c^\text{eff}=\frac15$. The other quantum dimension $d_W=\frac{1-\sqrt5}2$ can be studied in exactly the same way. Therefore, we do not repeat the details. What one has to know is that the allowed topological twists are
\[ \theta_W=e^{\pm\frac{2\pi i}5}. \]
The central charges are thus given as follows:
\begin{table}[H]
\begin{center}
\begin{tabular}{c|c|c}
$D_\text{Fibonacci}\backslash\theta_W$&$e^{+\frac{2\pi i}5}$&$e^{-\frac{2\pi i}5}$\\\hline
$+\sqrt{\frac{5-\sqrt5}2}$&$-\frac{11}{10}$&$-\frac{19}{10}$\\
$-\sqrt{\frac{5-\sqrt5}2}$&$\frac{29}{10}$&$\frac{21}{10}$
\end{tabular},
\end{center}
\caption{Central charges (mod $8$) of the rank six MTC with $T_1$}\label{c49T1minus}
\end{table}
\hspace{-18pt}and
\begin{table}[H]
\begin{center}
\begin{tabular}{c|c|c}
$D_\text{Fibonacci}\backslash\theta_W$&$e^{+\frac{2\pi i}5}$&$e^{-\frac{2\pi i}5}$\\\hline
$+\sqrt{\frac{5-\sqrt5}2}$&$\frac{29}{10}$&$\frac{21}{10}$\\
$-\sqrt{\frac{5-\sqrt5}2}$&$-\frac{11}{10}$&$-\frac{19}{10}$
\end{tabular}.
\end{center}
\caption{Central charges (mod $8$) of the rank six MTC with $T_2$}\label{c49T2minus}
\end{table}

After a routine exercise, one finds no solution. To conclude, this scenario --- the rank six MTC with $SU(2)_3/\mbb Z_2\times SU(2)_2$ realization --- has a unique solution, $c^\text{eff}=\frac15$.

\end{document}